\documentclass[epj]{svjour}

\usepackage{amsmath}
\usepackage{amssymb}
\usepackage{graphicx}

\begin{document}

\title{Local Detection of Entanglement}

\author{Gustavo Rigolin\inst{1} \and C. O. Escobar\inst{2}}
\institute{Departamento de F\'{\i}sica da Mat\'eria Condensada,
Instituto de F\'{\i}sica Gleb Wataghin, Universidade Estadual de
Campinas, C.P. 6165, cep 13084-971, Campinas, S\~ao Paulo, Brazil
\and Departamento de Raios C\'osmicos e Cronologia, Instituto de
F\'{\i}sica Gleb Wataghin, Universidade Estadual de Campinas, C.P.
6165, cep 13084-971, Campinas, S\~ao Paulo, Brazil}
\date{Received: date / Revised version: date}
\abstract{We construct an explicit model where it can be
established if a two mode pure Gaussian system is entangled or not
by acting only on one of the parts that constitute the system.
Measuring the dispersion in momentum and the time evolution of the
dispersion in position of one particle we can tell if entanglement
is present as well as the degree of entanglement of the system.
\PACS{
      {03.67.-a}{Quantum information}   \and
      {03.65Ud}{Entanglement and quantum nonlocality}
     } % end of PACS codes
} %end of abstract

\maketitle

\section{Introduction}

One of the most intriguing features of Quantum Mechanics (QM) is
entanglement and in the early times of QM it was recognized by
Erwin Schr\"{o}dinger \cite{schroedinger} and by Einstein,
Podolsky and Rosen \cite{epr}. Later, John Bell \cite{bell} showed
that the non-local aspect of entanglement is experimentally
testable through his famous inequality.

In recent years the interest in entanglement has increased considerably.
First because it is a fundamental tool in Quantum Information Theory and a
consistent characterization of its theoretical properties is needed.
Second because the present stage of technology permits us to perform some
experimental manipulations with it such as Quantum Teleportation
\cite{bennett93} and Quantum Cryptography \cite{bennett92,eckert}.

In the study of the properties of entanglement Peres \cite{peres} and the
Horodecki family \cite{h3} have derived a necessary and sufficient condition
for the separability of $2 \times 2$ and $2 \times 3$ systems. Some years
later Simon \cite{simon} and Duan {\em et al} \cite{duan} have obtained a
necessary and sufficient condition for the separability of two-party Gaussian
states. Given that the state is non-separable we should have a measure of the
degree of this inseparability. There are at least three distinct measures of
entanglement: the entanglement of formation \cite{bennettetal96c}, the
distillable entanglement \cite{bennettetal96a} and the relative entropy of
entanglement \cite{vedraletal97a,vedraletal97b}. In any calculations done with
these three measures of entanglement and the two criteria for
separability we must use the total density matrix of the bipartite system.
That is, given the density matrix that describes the whole bipartite system
we can determine if the system is separable or not and its degree of
entanglement.

For pure states it is well known that the knowledge of the whole reduced
density matrix allows us to decide whether or not the system is entangled.
If the reduced density matrix is pure ($\text{Tr}(\rho_{1}^{2})=1$) the system
is separable and entangled if it is not pure ($\text{Tr}(\rho_{1}^{2}) < 1$).
This article aims to show that we need not know the whole reduced density matrix
$\rho_{1}$ (or equivalently $\rho_{2}$) of a bipartite pure system to deduce if
it is entangled or not. Here we show that only the diagonal elements of the reduced
density matrix, even if it is written in a representation where it
is \textit{not diagonal}, are sufficient to detect entanglement.

To explicitly demonstrate this we construct two paradigmatic cases: a non-entangled
two particle Gaussian wave function in configuration space and an entangled two
particle Gaussian wave function. We then let both systems freely evolve in time.
We show that when studying an individual particle of each case we get different
results for the time evolution of the dispersion in position. This fact allows
us to tell if we are working with a non-entangled or an entangled bipartite Gaussian
wave function, and in the case of an entangled system we can also extract from this
evolution the degree of entanglement.

Some aspects of this approach are similar to the total wave function reconstruction
shown in Refs. \cite{leo,richter,ch}. Here, however,  we do not need to reconstruct
 the \textit{whole} wave function describing the two particles. We only need two
 elements of the reduced wave function describing a single particle, i. e., its
 dispersion in position and in momentum.  We should also mention that by employing
 the powerful techniques given in Refs. \cite{leo,richter} it might be possible
 to generalize the following approach to the case of mixed Gaussian states.

\section{The non-entangled bipartite system}

Consider a normalized one-dimensional separable two particle Gaussian wave
function where we assume, with no loss of generality, that the two particles
have the same mass $m$ but can in principle be distinguished from each other:

\begin{equation}
\psi(x_{1},x_{2},t) = \psi_{1}(x_{1},t) \otimes \psi_{2}(x_{2},t),
\label{geralemx}
\end{equation}
where
\begin{equation}
\psi_{1}(x_{1},t)=\int_{}^{} f(k_{1}) e^{i[k_{1}x_{1} - \omega(k_{1})t]}
dk_{1}, \label{dois}
\end{equation}
\begin{equation}
\psi_{2}(x_{2},t)=\int_{}^{} f(-k_{2}) e^{i[k_{2}x_{2} - \omega(k_{2})t]}
dk_{2}. \label{tres}
\end{equation}
Here $\omega(k)=\frac{\hbar k^{2}}{2 m}$ is the dispersion
relation for a free particle and $f(k_{1})$, $f(-k_{2})$
represents the fact that we have Gaussian particles moving in
opposite directions \cite{cohen}:
\begin{equation}
f(k)= \frac{\sqrt{a}}{(2\pi)^{3/4}}e^{-\frac{a^{2}}{4}(k-k_{c})^{2}}.
\label{fdk}
\end{equation}
In Eq.~(\ref{fdk}), $a$ represents the dispersion of the Gaussian wave packet
centered in $k_{c}$ and the factor that multiplies the exponential is the
normalization constant.

Integrating in $k_{1}$ and $k_{2}$, then  multiplying
$\psi(x_{1},x_{2},t)$ by its complex conjugate and finally
integrating in $x_{2}$ we get the probability density of particle
$1$ at time $t$ \cite{cohen}:

\begin{equation}
|\varphi(x_{1},t)|^{2}=\sqrt{\frac{2}{\pi a^{2}}} \frac{1}{\sqrt{1+F(t)}}
\exp \left[ -\frac{2}{a^{2}} \frac{(x_{1}-v_{c} t)^{2}}{1+F(t)} \right],
\label{gaussx1}
\end{equation}
where $F(t)$ and $v_{c}$ are given by:
\begin{eqnarray}
F(t) = F(t,a) = \frac{4\hbar^{2} t^{2}}{m^{2}a^{4}},& \;\;\;\;\; & v_{c} =
\frac{\hbar k_{c}}{m}.
\end{eqnarray}

With Eq.~(\ref{gaussx1}) we can calculate the dispersion
$\Delta x_{1}=\sqrt{\left<x_{1}^{2}\right>-\left<x_{1}\right>^{2}}$ of
the position of particle $1$:

\begin{equation}
\Delta x_{1}(t)=\frac{a}{2}\sqrt{1+F(t)}. \label{delx1ne}
\end{equation}
We can also obtain the dispersion of the momentum of particle $1$ if we take
the Fourier transform of Eq.~(\ref{geralemx}). Then  multiplying the result
by its complex conjugate and integrating in $k_{2}$ we obtain:
\begin{equation}
|\tilde{\varphi}(k_{1},t)|^{2}=\sqrt{\frac{a^{2}}{2\pi}}
\exp \left[ -\frac{a^{2}}{2} (k_{1}-k_{c})^{2} \right].
\label{gaussp1}
\end{equation}
Using Eq.~(\ref{gaussp1}) and the fact that $p_{1}=\hbar k_{1}$ we easily get:
\begin{equation}
\Delta p_{1}(t)=\frac{\hbar}{a}. \label{delp1ne}
\end{equation}
As expected for a free particle the dispersion in momentum is constant in time.

\section{The entangled bipartite system}

Let us now construct a normalized one-dimensional entangled two particle
Gaussian wave function where we assume again, with no loss of generality,
that the two particles have the same mass $m$ but can in principle be
distinguished from each other.

\begin{eqnarray}
\Psi(x_{1},x_{2},t=0) & = & \int_{}^{}  dk_{1} dk_{2} f(k_{1},k_{2}) \nonumber \\
& & \times \psi_{1}(x_{1},0) \otimes \psi_{2}(x_{2},0). \label{ent}
\end{eqnarray}
Here $ \psi_{1}(x_{1},0)$ and $ \psi_{2}(x_{2},0)$ are given by:
\begin{equation}
\psi_{1}(x_{1},0)= e^{ik_{1} x_{1}} e^{\frac{-x_{1}^{2}}{a^{2}}}, \label{g1}
\end{equation}
\begin{equation}
\psi_{2}(x_{2},0)= e^{ik_{2} x_{2}} e^{\frac{-x_{2}^{2}}{a^{2}}}. \label{g2}
\end{equation}
Eq.~(\ref{ent}) is a superposition of bipartite Gaussian wave packets
centered in $k_{1}$ and $k_{2}$ where $f(k_{1},k_{2})$ $=$ $g(k_{1},k_{2})$
$\delta(k_{1}+k_{2})$ are the expansion
coefficients and $\delta(k_{1}+k_{2})$ is a restriction which entangles the
system. This delta function can be viewed as the requirement for the
conservation of momentum in the center of mass frame, that is, we superpose
bipartite Gaussian wave packets where each party moves in opposite
directions centered at the same momentum.
Eqs.~(\ref{g1}) and (\ref{g2}) are proportional to Eqs.~(\ref{dois}) and
(\ref{tres}) where we integrate for $t=0$ and substitute
$k_{c}$ by $k_{1}$ and $k_{2}$ respectively.
By using the delta function Eq.~(\ref{ent}) can be rewritten as:
\begin{eqnarray}
\Psi(x_{1},x_{2},0) & = & \int_{}^{}dk_{1}g(k_{1}) \nonumber \\
& & \times \left( e^{ik_{1} x_{1}} e^{\frac{-x_{1}^{2}}{a^{2}}} \right)
\left( e^{-ik_{1} x_{2}} e^{\frac{-x_{2}^{2}}{a^{2}}} \right). \label{quaseepr}
\end{eqnarray}
Eq.~(\ref{quaseepr}) clearly shows that $\delta(k_{1}+k_{2})$
entangles our system. Because $\Psi(x_{1},x_{2},0)$ cannot be
written as a simple tensor product of a wave function belonging to
particle $1$ and another belonging to particle $2$ we now deal
with a non-separable wave function. Only if $g(k_{1})$ is another
delta function we can disentangle the system and recover
Eq.~(\ref{geralemx}). In Eq.~(\ref{quaseepr}) $g(k_{1})$ is chosen
to be a Gaussian distribution centered in $k_{c}$:
\begin{equation}
g(k_{1})= \sqrt{\frac{2}{\pi a^{2}}}f_{2}^{\frac{1}{4}}\frac{(b/2)}{\sqrt{\pi}}
\exp \left[ -(b/2)^{2} (k_{1}-k_{c})^{2}\right], \label{coeficiente}
\end{equation}
where $b$ is a new parameter that measures the degree of entanglement as
explained below and $f_{n}=1+n \frac{a^{2}}{b^{2}}$, $n=1,2$.
We can see that when $b \rightarrow \infty$ the function
$\frac{(b/2)}{\sqrt{\pi}}\exp \left[ -(b/2)^{2} (k_{1}-k_{c})^{2}\right]$
$\rightarrow$ $\delta(k_{1}-k_{c})$ \cite{arfken} and $f_{2}\rightarrow 1$,
showing that entanglement has disappeared. This can be seen doing a
straightforward calculation using Eqs.~(\ref{coeficiente}) and (\ref{quaseepr}):
\begin{eqnarray}
\lim_{b \rightarrow \infty} \Psi(x_{1},x_{2},0) & = &
\left[ \left(\frac{2}{\pi a^{2}}\right)^{1/4} e^{ik_{c} x_{1}}
e^{\frac{-x_{1}^{2}}{a^{2}}} \right] \nonumber \\
& \otimes & \left[ \left( \frac{2}{\pi a^{2}}\right)^{1/4} e^{-ik_{c} x_{2}}
e^{\frac{-x_{2}^{2}}{a^{2}}} \right]. \label{amesmaqueum}
\end{eqnarray}
Eq.~(\ref{amesmaqueum}) is identical to Eq.~(\ref{geralemx}) if we calculate
the integrals in Eqs.~(\ref{dois}) and (\ref{tres}).
Furthermore, it can be shown that if $b \rightarrow 0$ and $a \rightarrow \infty$
Eq.~(\ref{quaseepr}) is the EPR state with $x_{0}=0$ \cite{englert}. As stated
in ref. \cite{englert}, Eq.~(\ref{quaseepr}) can be viewed as a generalized version
of the EPR wave function.
These two facts suggest that $b$ should be considered as a measure of the
degree of entanglement, where $b \rightarrow \infty$ represents no entanglement and
$b \rightarrow 0$ represents the maximally entangled state.

Doing the integral in Eq.~(\ref{quaseepr}) we get the normalized bipartite
Gaussian wave function at $t=0$:
\begin{eqnarray}
\Psi(x_{1},x_{2},0) &= & \sqrt{\frac{2}{\pi a^{2}}}f_{2}^{\frac{1}{4}}
e^{ik_{c} (x_{1}-x_{2})} \nonumber \\
& & \times \exp \left[-\frac{f_{1}}{a^{2}}(x_{1}^2+x_{2}^{2})+
\frac{2}{b^{2}}x_{1}x_{2} \right]. \label{16}
\end{eqnarray}
It is interesting to note that Eq.~(\ref{16}) represents a non-separable
(entangled) state due to the term $\exp \left[ \frac{2}{b^{2}}x_{1}x_{2} \right]$.
If $b \rightarrow \infty$ this term tends to $1$ and we obtain Eq.~(\ref{geralemx})
as a limiting case of Eq.~(\ref{16}). In other words,
when $b \rightarrow \infty$ we have Eq.~(\ref{geralemx}),
a separable, non-entangled state, and for any other value of $b$
we have Eq.~(\ref{16}), a non-separable, entangled state.

In order to make rigorous that $b$ furnishes the degree of
entanglement of the state given by Eq.~(\ref{16}) and that only
when $b \rightarrow \infty$ we have a disentangled system we first
calculate its correlation matrix (CM) and apply the Simon separability
criterion \cite{simon}, which shows that the bipartite Gaussian system
is separable iff $b \rightarrow \infty$. After applying the Simon criterion,
we make a local symplectic transformation in the CM to put it in its
standard form \cite{simon,duan} and then calculate its entanglement of
formation (EoF) \cite{werner}, which is a monotonically decreasing
function of the parameter $b$, proving that the higher $b$ the less
entangled is the state. As we deal with a pure state, we note that
we can calculate the von Neumann entropy of the reduced density matrix
to obtain the entanglement of this system. However, we prefer using the
EoF as given in Ref. \cite{werner} since, in the particular case of
symmetric bipartite Gaussian states, it is more straightforward than the usual
procedure for pure states.

The CM completely specify a two mode Gaussian state and it is a
$4 \times 4$ matrix, which has the following elements \cite{simon,duan}:
\begin{equation}
\gamma_{ij}=\text{Tr}\left[(R_{i}R_{j}+R_{j}R_{i})\rho\right]-2
\text{Tr}[R_{i}\rho]\text{Tr}[R_{j}\rho],
\end{equation}
where $R=(X_{1},P_{1},X_{2},P_{2})^{T}$ and $R_{j}$ are the
position and momentum operators of the two particles. Doing the
calculations we get the following CM:
\begin{equation}
\gamma =
\left(
\begin{array}{cc}
A & C \\
C^{T} & A
\end{array}
\right), \label{gama}
\end{equation}
where
\begin{equation}
 \begin{array}{ccc}
   A = \left(
    \begin{array}{cc}
    \frac{a^{2}f_{1}}{2f_{2}} & 0 \\
    0 & \frac{2\hbar^{2}f_{1}}{a^{2}}
    \end{array}
    \right), &
   C =  \left(
    \begin{array}{cc}
    \frac{a^{4}}{2b^{2}f_{2}} & 0 \\
    0 & -\frac{2\hbar^{2}}{b^{2}}
    \end{array}
    \right).
 \end{array}
\end{equation}
The Simon separability criterion says that the above CM represents a
non-entangled system iff \cite{nota1}:
\begin{eqnarray}
I & = & \text{det}A\text{det}B + \left(  \hbar^{2} - \left|
\text{det}C \right| \right)^2 - \text{Tr}\{ AJCJBJC^{T}J  \} \nonumber \\
& & - \hbar^{2}(\text{det}A + \text{det}B) \geq 0, \label{criterio}
\end{eqnarray}
where $J=
\left(
\begin{array}{cc}
0 & 1 \\
-1 & 0
\end{array}
\right)$.
But a simple calculation shows that the rhs of Eq.~(\ref{criterio}) is:
\begin{equation}
I = - 4\hbar^{4}\frac{a^{4}}{b^{4}}\frac{1}{f_{2}}. \label{crit}
\end{equation}
Hence, $I < 0$ except when $b \rightarrow \infty$, proving that for
any other value of $b$ we have an entangled state.

\begin{figure}[!ht]
\begin{center}
\includegraphics[width=5cm]{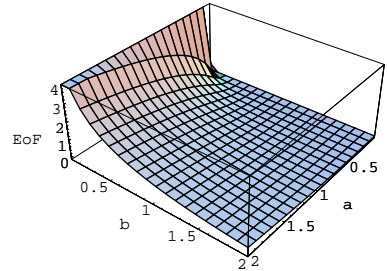}
\end{center}
\caption{ \small{Entanglement of formation, Eq.~(\ref{efs}), for the
symmetric Gaussian state given by Eq.~(\ref{16}), as function of the
parameters $a$ and $b$, where we have put $\hbar =1$. We clearly see the
EoF increasing as $b \rightarrow 0$  and decreasing as  $a \rightarrow 0$.}}
\label{eof1}
\end{figure}

\begin{figure}[!ht]
\begin{center}
\includegraphics[width=5cm]{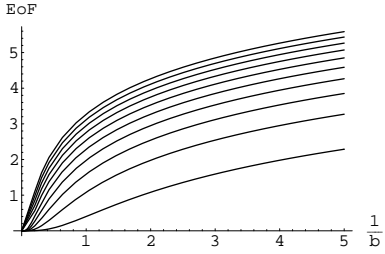}
\end{center}
\caption{\small{Entanglement of formation, Eq.~(\ref{efs}), as a function
of $1/b$ for ten values of the parameter $a$. From bottom to top the
parameter $a$ varies from $1$ to $10$ in increments of one unit. We have
set $\hbar=1$. We clearly see that the EoF increases as $b$ decreases and
that for a given b, the higher $a$, the greater the EoF.}}
\label{eof2}
\end{figure}

We now make the following local symplectic transformation
$S$ $=$ $\text{diag}$ $(s,$ $s^{-1},$ $s,$ $s^{-1})$, where
$s$ $=$ $\left(4\hbar^{2}f_{2}/a^{4}\right)^{1/4}$. This brings
$\gamma$ to its standard form $\gamma_{0} = S\gamma S^{T}$
\cite{simon,duan}:
\begin{equation}
\gamma_{0}=
\left(
\begin{array}{cccc}
n & 0 & k_{x} & 0 \\
0 & n & 0 & -k_{p} \\
k_{x} & 0 & n & 0 \\
0 & -k_{p} & 0 & n
\end{array}
\right), \label{standard}
\end{equation}
where $n =\hbar f_{1}/\sqrt{f_{2}}$ and $k_{x}=k_{p}=\hbar a^{2}/(b^{2}\sqrt{f_{2}})$.
This is a symmetric Gaussian system and Giedke \textit{et al.}
\cite{werner} have shown that the EoF for this state is:
\begin{equation}
EoF(\Psi) = f\left[\sqrt{(n-k_{x})(n-k_{p})}\right], \label{efs}
\end{equation}
where,
\begin{equation}
f(\delta)=c_{+}(\delta)\,\log_{2}[c_{+}(\delta)]-c_{-}(\delta)\,
\log_{2}[c_{-}(\delta)].
\end{equation}
Here $c_{\pm}(\delta)=(\delta^{-1/2}\pm \delta^{1/2})^2/4$.
Analyzing the behavior of the EoF given by Eq.~(\ref{efs}) we
clearly see that it is a decreasing function of the parameter $b$
(Figs.~\ref{eof1} and \ref{eof2}).


Working in the Heisenberg picture we easily obtain for a free evolution,
\begin{equation}
\Delta x_{1}(t)=\frac{a}{2} \sqrt{\frac{f_{1}}{f_{2}}[1+f_{2}F(t)]}, \label{delx1ent}
\end{equation}
\begin{equation}
\Delta p_{1}(t)=\frac{\hbar}{a}\sqrt{f_{1}}. \label{delp1e}
\end{equation}
Again, due to the free evolution of particle $1$ the dispersion in momentum
does not vary in time.

We should mention that the formal solution of the Heisenberg equations
of motion for the observables $x_{1}(t)$, $ x_{1}^{2}(t)$, $p_{1}(t)$,
and $p_{1}^{2}(t)$ are identical for the entangled and non-entangled case.
Only when we take the mean values $\langle x_{1}(t)\rangle$,
$\langle x_{1}^{2}(t)\rangle$, $\langle p_{1}(t)\rangle$, and
$\langle p_{1}^{2}(t)\rangle$ we obtain different quantities.
This is due to the fact that we have different initial wave functions.
In other words, entanglement manifests itself furnishing different
initial conditions for the Heisenberg equations of motion, which imply
different evolutions for the dispersions.

\section{The measuring protocol}

As we have all tools now, that is, all the dispersions in position and in
momentum for the entangled and non-entangled case, we develop a measurement
procedure to be used in an ensemble of two particle Gaussian systems
which allows us to locally decide whether or not the particles are entangled. From
now on we assume $\hbar=m=1$ for simplicity.

Let Bob be our physicist who receives one of the particles of the bipartite
Gaussian system produced by Alice. Bob knows, because Alice has told him,
that all the particles he receives are either entangled or non-entangled
Gaussian wave packets, according to the two constructions explained above.
There are no other possibilities. Alice produces many pairs at once.
And continues to produce many pairs at once for different times.
Of course Bob does not know the values of the parameters $a$ and $b$ used
by Alice. But Bob is curious enough and wants to know whether his particles
are entangled or not. Bob cannot use any further classical communication,
he can act only locally on his particles and he is able only to measure
the dispersions in position and in momentum of his particles, that is,
the diagonal elements of the system reduced density matrix. He proceeds as follows:

First he measures, using a sub-ensemble, the dispersion in momentum of his
wave packets. He obtains $\Delta p_{1}=u$. He does not know yet whether
Eq.~(\ref{delp1ne}) or Eq.~(\ref{delp1e}) represents what he measures.
However he knows that it must be one of these two possibilities, which imply
only two possible time evolution for the dispersion in the position of his
particles.

If his particles are not entangled and Bob uses in Eq.
(\ref{delx1ne}) the fact that $\Delta p_{1}(t)=u=\frac{1}{a}$ he
gets:
\begin{equation}
\Delta x_{1}(t)= \frac{1}{2 u}\sqrt{1+4u^{4}t^{2}}. \label{the1}
\end{equation}
But if Bob's particles are entangled and now he uses the fact that
$\Delta p_{1}(t)=u=\frac{\sqrt{f_{1}}}{a}$, Eq.~(\ref{delx1ent}) becomes:
\begin{equation}
\Delta x_{1}(t)= \frac{1}{2 u}\sqrt{\frac{u^{4}b^{4}}{u^4b^{4}-1}+4u^{4}t^{2}}.
\label{the2}
\end{equation}
Looking at Eqs.~(\ref{the1}) and (\ref{the2}) we see that if Bob knows at
what time Alice has begun to produce the pairs he is able to discover,
with only one measurement of $\Delta x_{1}$, whether his particles are
entangled with Alice's or not. The reason for this is simple: Let us suppose,
with no loss of generality, that Alice begins to produce the pairs of
particles at $t=0$. Measuring the dispersion in position for a given time $t$
Bob obtains $\Delta x_{1}(t)$. Remembering that Bob also knows the value
of $u$, he can calculate, using Eq.~(\ref{the1}), the value of $\Delta x_{1}(t)$.
If this calculated value of the dispersion agrees with the measured one,
Bob has the non-entangled case. If these values of $\Delta x_{1}$ are different,
Bob has entanglement. In this last case, using Eq.~(\ref{the2}) Bob can
obtain the parameter $b$. For any $t$ Bob can use this procedure. Bob
sees two distinct curves for the
time evolution of $\Delta x_{1}(t)$, whether his particles are entangled or
not. See Fig.~$1$ below:

\begin{figure}[!ht]
\begin{center}
\includegraphics[width=5cm]{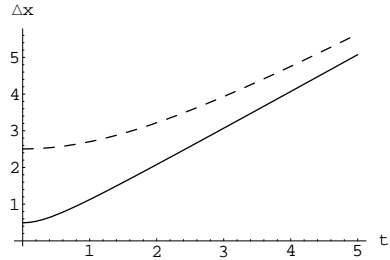}
\end{center}
\caption{\small{The dashed curve is the time evolution of the
dispersion in position for an entangled Gaussian wave packet while the solid
curve represents the non-entangled case. We have chosen $b=1$ and $u=1.01$.}}
\label{label}
\end{figure}

Analyzing Eq.~(\ref{the2}) we see that for it to be valid for all $t \geq 0$
we must have for the entangled case:
\begin{equation}
u b > 1. \label{desigualdade}
\end{equation}
It is worth noting that asymptotically Eq.~(\ref{the1}) and (\ref{the2}) are
the same. Therefore, in order for Bob to correctly distinguish between the
two cases he should make his measurements for times smaller than a critical
time $t_{c}$, which is defined to be the time where the time independent term
inside the square root of Eq.~(\ref{the2}) is of the order of the $t^{2}$ term:
\begin{equation}
t_{c} \approx \frac{b^{2}}{2\sqrt{u^{4}b^{4}-1}}. \label{tempocritico}
\end{equation}
We can increase $t_{c}$ making $ub \rightarrow 1$. This might seem as a
limitation of our procedure but as Alice sends a classical message to Bob
defining the origin of time, Bob can start making measurements as early as
possible.

Now let us make things harder to Bob. We assume from now on that Bob
does not know when and where Alice has begun to produce the pairs.
This fact means that Bob cannot use the previous procedure to answer
whether or not his particles are entangled with Alice's. The previous
protocol fails because Bob does not know what time $t$ he should use
to calculate $\Delta x_{1}(t)$, which would have allowed him to
compare this calculated value with the measured $\Delta x_{1}(t)$.

We first prove why a single measurement at time $t$ is not enough
for Bob to tell whether his particles are entangled or not.  We
are now assuming that he does not know when Alice has begun to
produce the particles. The proof is achieved showing that the
diagonal elements of the reduced density matrix (in position
and in momentum representation) of the non-entangled system
can be made identical to the diagonal elements of the reduced
density matrix of the entangled system for $t=0$. (The same
reasoning applies to any $t$, but for $t=0$ the calculations
are much simpler and we do not lose in generality).

For $t=0$ the diagonal elements of the reduced density matrix of the
entangled system in momentum representation is
\begin{eqnarray}
\varrho_{1} (k_{1}) & = &
\int_{}^{}\left<k_{1},k_{2} \right|\Psi\left>\right<\Psi\left|k_{1},k_{2}\right> dk_{2}
\nonumber \\
& = & \sqrt{\frac{a^{2}}{2\pi f_{1}}} \exp \left[ -\frac{a^2}{2f_{1}}(k_{1}-k_{c})^{2}
\right].
\label{ematmom}
\end{eqnarray}
For any $t$, the diagonal elements of the reduced density matrix of the
non-entangled system in the momentum representation, according to
Eq.~(\ref{gaussp1}), reads:
\begin{eqnarray}
\rho_{1} (k_{1},t) & = & \int_{}^{}\left<k_{1},k_{2}
\right|\psi\left>\right<\psi\left|k_{1},k_{2}\right> dk_{2} \nonumber \\
& = & \sqrt{\frac{a'^{2}}{2\pi}}
\exp \left[ -\frac{a'^{2}}{2} (k_{1}-k_{c})^{2} \right]. \label{nematmom}
\end{eqnarray}
If we want identical diagonal elements of the reduced density
matrices we must impose that:
\begin{equation}
 a'=\frac{a}{\sqrt{f_{1}}}=\frac{1}{u}. \label{definicao}
\end{equation}

\begin{figure}[!ht]
\begin{center}
\includegraphics[width=5cm]{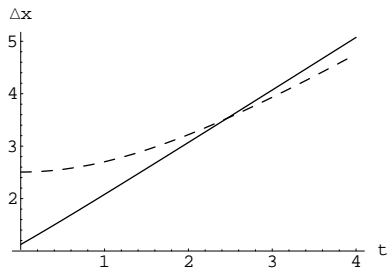}
\end{center}
\caption{\small{The dashed curve is the time evolution of the
dispersion in position for an entangled Gaussian wave packet produced $1$
unity of time after the production of the non-entangled case, which is represented
by the solid curve. The curves intercept each other for $t \approx 2.46$. If Bob
measures $\Delta x_{1}$ for this time, he cannot distinguish between the two ways
in which Alice can produce the pairs of particles. For any other point of the
dashed curve we can find a solid one that crosses it. Therefore, Bob cannot
distinguish how his particles were produced if he measures $\Delta x_{1}$ only
once. Here we have chosen $b=1$ and $u=1.01$.}}
\label{label2}
\end{figure}

The diagonal elements of the reduced density matrix for $t=0$ of the
entangled system written in the position representation is:
\begin{eqnarray}
\varrho_{1} (x_{1}) & = &
\int_{}^{}\left<x_{1},x_{2} \right|\Psi\left>\right<\Psi\left|x_{1},x_{2}\right> dx_{2}
\nonumber \\
& = & \sqrt{\frac{2f_{2}}{\pi a^{2}f_{1}}} \exp
\left[ -\frac{2f_{2}}{a^{2}f_{1}} x_{1}^{2} \right].
\label{ematpos}
\end{eqnarray}
As stated in Eq.~(\ref{gaussx1}), the diagonal elements of the reduced
density matrix for any $t$ of the non-entangled system in position representation is:
\begin{eqnarray}
\rho_{1} (x_{1}) & = &
\int_{}^{}\left<x_{1},x_{2} \right|\psi\left>\right<\psi\left|x_{1},x_{2}\right> dx_{2}
\nonumber \\
& = & \sqrt{\frac{2}{\pi a'^{2}}} \frac{1}{\sqrt{1+F(t,a')}} \nonumber \\
& & \times \exp \left[ -\frac{2}{a'^{2}} \frac{(x_{1}-v_{c} t)^{2}}{1+F(t,a')} \right].
\nonumber \\
& & \; \label{nematpos}
\end{eqnarray}
If we want Eqs.~(\ref{ematpos}) and (\ref{nematpos}) giving the same statistical
predictions we must have:
\begin{equation}
\frac{2f_{2}}{a^{2}f_{1}} = \frac{2}{a'^{2}}\frac{1}{1+F(t,a')}. \label{condicao}
\end{equation}
Eq.~(\ref{condicao}) is a restriction which forces the two density
matrix to give the same dispersion in position. (We do not need to
bother with the first order moment of these Gaus\-si\-an functions
because a translation of the $x_{1}$-axis sets it to zero.) If we
use Eqs.~(\ref{definicao}) and (\ref{condicao}) and the fact that
$f_{n} = 1 + n \frac{a^{2}}{b^{2}}$ we arrive at the following
condition:
\begin{equation}
t=\frac{1}{2u^{2}}\frac{1}{\sqrt{u^{4}b^{4}-1}} \label{tempo}
\end{equation}
Eq.~(\ref{tempo}) says that for only, and only one time $t$, the diagonal
elements of the reduced density matrices, one obtained from the entangled
system and the other one obtained from the non-entangled system, furnish the
same statistical predictions. This implies that single measurements of the
dispersion in momentum and in position of particle $1$ do not tell us unequivocally
whether we are dealing with a non-entangled or entangled Gaussian bipartite system.
(Unless, of course, we know when Alice has begun to produce the pairs.)
See Fig.~$\ref{label2}$.

To circumvent the limitation of the previous protocol Bob may apply the
following one, which explicitly uses the difference in time evolution of the
two systems:

Bob again initially measures the dispersion in momentum of his particles
($\Delta p_{1}=u$). As he does not know when and where Alice begins to produce
the pairs of Gaussian particles, the time evolution of the dispersions in position
for the non-entangled and entangled systems are:
\begin{equation}
\Delta x_{1}(t)= \frac{1}{2 u}\sqrt{1+4u^{4}(t+t_{0})^{2}}. \label{the3}
\end{equation}
\begin{equation}
\Delta x_{1}(t)= \frac{1}{2 u}\sqrt{\frac{u^{4}b^{4}}{u^4b^{4}-1}+4u^{4}(t+t_{0})^{2}}.
\label{the4}
\end{equation}
Here $t_{0}$ is the time elapsed from the production of the pair
by Alice until Bob makes his first set of measurements. Bob now
makes several measurements of the dispersion in position for
different times $t$. With these measurements he obtains the
following set of points:
\begin{displaymath}
\{(\Delta x_{1}(0),0), (\Delta x_{1}(t_{1}),t_{1}), \ldots,
(\Delta x_{1}(t_{n}), t_{n})\}.
\end{displaymath}
He makes as many measurements as possible. With the $n$ pairs of
points above he fits the following curve, where $\alpha$ and
$\beta$ are the free parameters and $u$ is already known:

\begin{equation}
\Delta x_{1}(t)= \frac{1}{2 u}\sqrt{\alpha + 4u^{4}(t+\beta)^{2}}. \label{fit}
\end{equation}
Looking at Eqs.~(\ref{fit}), (\ref{the3}), and (\ref{the4}) we see that
if the parameter $\alpha=1$ Bob is dealing with  non-entangled Gaussian
functions, but if $\alpha \neq 1$ Bob deals with entangled particles. And
using $\alpha$ Bob can calculate the value of the degree of entanglement $b$.
Just for completeness we mention that $\beta$ furnishes the time $t_{0}$.
For this protocol to be optimal, Bob should begin his measurements as soon
as possible since,  asymptotically in time, Eqs.~(\ref{the3}) and (\ref{the4})
are seen to become identical.
\section{Conclusion}

We have shown an explicit model using two particle Gaus\-si\-an
systems where we can decide if we are dealing with non-entangled
or entangled pairs acting only on one of the particles and
measuring only the diagonal elements of its reduced density
matrix. Measuring the dispersion in momentum and then the time
evolution of the dispersion in position of one member of the pair
it is possible to discern between the entangled and non-entangled
cases. It is also possible with this procedure to determine the
degree of the entanglement of the system. The above model suggests
that just one part of the whole system can furnish more
information about the degree of the entanglement of the system
than we had imagined.

Finally it is important to note that the presented measurement protocol uses
the time evolution of the diagonal elements of the reduced density matrix to
determine whether or not we have entanglement. This fact shows that we may have
a new tool to analyze the properties of entangled systems, i. e., the dynamical
evolution of entangled states. So far all the methods used to study the properties of
entangled systems have not employed the dynamics of the system. We are hopeful
that studying the dynamics of entangled systems will help us to deepen our
understanding of entanglement and possibly it will unravel new features of
entanglement not yet explored.

\section*{Acknowledgments}
This work was supported by Funda\c{c}\~ao de Amparo \`a
Pes\-qui\-sa do Estado de S\~ao Pau\-lo (FAPESP) and Conselho
Nacional de Desenvolvimento Cient\'{\i}fico e Tecnol\'ogi\-co
(C\-N\-P\-q). We thank Dr. L\'ea F. dos Santos for her careful
reading of the manuscript and useful discussions.


\begin{thebibliography}{25}
\bibitem{nota1} To recover the Simon separability criterion as it was presented in
Simon's paper \cite{simon} we should substitute $\hbar^2$ in Eq.~(\ref{criterio})
by $\hbar^{2}/4$. This difference is due to the fact that Simon's CM is defined as
the half of our CM $\gamma$. We use the same notation as \cite{duan}.
\end{thebibliography}
\end{document}